\documentclass[aps,pra,showpacs,twocolumn,letterpaper,nobibnotes]{revtex4-1}
\usepackage{amssymb}
\usepackage{amsmath}
\usepackage{graphicx}
\usepackage{url}
\newcommand\ket[1]{\ensuremath{|#1\rangle}}

\newcommand\mean[1]{\ensuremath{\left<#1\right>}}

\begin{document}
\title{Entangled collective-spin states of atomic ensembles under non-uniform atom-light interaction}
\author{Jiazhong Hu}
\author{Wenlan Chen}
\author{Zachary Vendeiro}
\author{Hao Zhang}
\author{Vladan Vuleti\'{c}}
\affiliation{
Department of Physics and Research Laboratory of Electronics, Massachusetts Institute of Technology,
Cambridge, Massachusetts 02139, USA}
\begin{abstract}
We consider the optical generation and verification of entanglement in atomic ensembles under non-uniform interaction between the ensemble and an optical mode. We show that for a wide range of parameters a system of non-uniformly coupled atomic spins can be described as an ensemble of uniformly coupled spins with a reduced effective atom-light coupling and a reduced effective atom number, with a reduction factor of order unity given by the ensemble-mode geometry. This description is valid even for complex entangled states with arbitrary phase-space distribution functions as long as the detection does not resolve single spins. Furthermore, we derive an analytic formula for the observable entanglement in the case, of relevance in practice, where the ensemble-mode coupling differs between state generation and measurement.
\end{abstract}
\maketitle
\section{Introduction}

In cavity quantum electrodynamics (cQED), an optical resonator enhances the interaction between atoms and light. A particularly interesting regime is reached when the back action of the atoms on the cavity and the back action of the cavity field on the atoms become appreciable. In this strong-coupling regime where the system can evolve reversibly and coherently, many interesting experiments can be realized \cite{kimble1995,kimble2005,reichel2007,thompson2011,reichel40atoms,SpinMeasurement,SpinDy}. For instance, it is possible to realize measurements beyond the standard quantum limit \cite{thompson2011,SpinMeasurement,SpinDy,10db} by preparing a particular class of entangled states, spin squeezed states. These states are typically prepared using a non-uniform light-atom interaction. Recently, a different entangled state of many atoms described by a negative-valued, doughnut-shaped Wigner function has been realized using the strong collective light-atom interaction in a standing-wave optical cavity with manifestly non-uniform atom-light coupling \cite{3000atoms}.

Most treatments of atom-light coupling \cite{squeeze,PRAHei,CavityAid,PRAtheory} consider the situation where the atoms and light are uniformly coupled. However, in real systems, this assumption is hardly ever fulfilled. For instance, when the atomic cloud is comparable to or larger than the waist of the light mode to which the atoms are coupled, it is necessary to take into account the inhomogeneity of the atom-light coupling caused by the mode profile. In general, when the light intensity is not uniform in the volume occupied by the atoms, non-uniform atom-light coupling occurs. While this could be remedied by using a larger beam, this is often undesirable, as it reduces the strength of the atom-light interaction \cite{TanjiSuzuki2011201}. More generally, the coupling is always non-uniform at some level, for instance, due to thermal motion of the atoms. The effect of inhomogeneous coupling is more severe for highly entangled states.

Theoretical work on non-uniformly coupled atom-light systems has focused on Gaussian states \cite{Kennedy,Klaus,Retamal} , where the atomic quasi probability function is described by a Gaussian function. However, it is not immediately obvious whether non-Gaussian entangled states \cite{PRAtheory,reichel40atoms,3000atoms,Strobel,Polzik1,Polzik2} can be generated and detected under non-uniform atom-light coupling.

\begin{figure}[htbp]
\begin{center}
\includegraphics[height=1.5in]{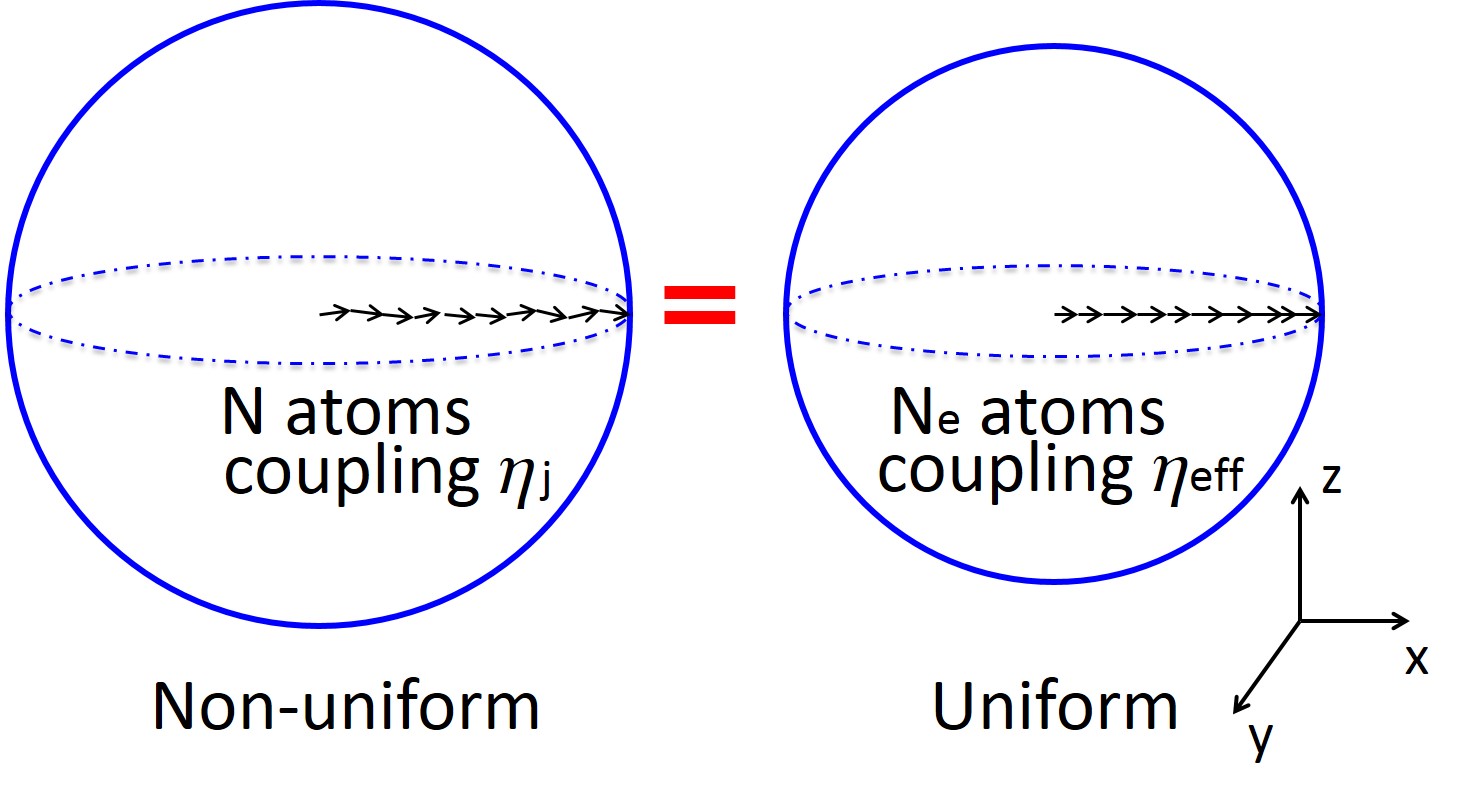}
\caption{Equivalence between a system of $N$ spins non-uniformly coupled to an optical mode and a slightly smaller uniformly coupled system of $N_e$ spins. The equivalence is valid when the inidvidual atomic spins are approximately aligned, $|\vec S|\approx Ns$. The uniformly coupled system consists of $N_e=N\mean{\eta_j}^2/\mean{\eta^2_j}$ effective atoms coupled with effective strength $\eta_{eff}=\mean{\eta^2_j}/\mean{\eta_j}$, where $\eta_j$ is the coupling strength for each atom $j$. Replacing the spin operators by effective spin operators (see text), within a wide range of parameters all dynamical properties are the same as those of the uniformly coupled system. }
\label{fig1}
\end{center}
\end{figure}

For uniform coupling, the collective spin degrees of freedom are well described by the total spin components $S_x$, $S_y$ and $S_z$, with the eigenstates of $S_z$ (or $S_x$, $S_y$) being the Dicke states \cite{Dicke}. The question then is whether similar collective operators can be found to describe the evolution and measurement of the collective spin under non-uniform coupling.

In this article, we prove that for a wide range of states the non-uniformly coupled system is equivalent to a (slightly smaller) uniformly coupled system when the atom number $N$ is large, and when the individual atomic spins are mostly aligned with each other, i.e. $|\vec S|\approx Ns$, where $s$ is the spin of a single atom. We show that under a wide range of conditions, we can simply replace the spin operator $S_x$, $S_y$, $S_z$ by appropriately defined effective spin operators $\tilde S_x$, $\tilde S_y$, $\tilde S_z$ to describe the system. The system dynamics are then the same as those of a uniformly coupled system. We also define effective Dicke states under non-uniform coupling, and generalize the concept of the effective atom number $N_e$ that was first introduced in Ref \cite{SpinMeasurement,SpinDy}, and that has been applied to several experiments \cite{10db,clock,PRAtheory}.

\section{The equivalence between uniform and non-uniform coupling}

To be specific, we consider here the quantum non-demolition interaction that is used for most experiments \cite{reichel40atoms,thompson2011,SpinMeasurement,10db,SpinDy,3000atoms,CavityAid} such as spin squeezing and entangled states generation. It has the form
\begin{equation}
H=\hbar\Omega \hat S_z \hat A.\label{eq1}
\end{equation}
Here, $\hat S_z=\sum_{j=1}^N \hat s^{(j)}_z$, where $\hat s^{(j)}_z$ is the spin operator along the $z$ axis of atom $j$, and $\hat A$ is any Hermitian operator of the light field.

A Hamiltonian of this form appears in a variety of situations. For instance, if $\hat A=\hat c^\dagger \hat c$ \cite{SpinMeasurement,SpinDy,thompson2011,10db}, which is the intensity operator of the light, where $\hat c$ is the annihilation operator for a photon in the electromagnetic mode of interest, then $H$ describes the shift of the cavity resonance frequency by the atoms, or equivalently, the light shift on the atoms by the intracavity field. If $\hat A=\hat J_z$ \cite{Polzik1,Polzik2,3000atoms,Kennedy,Klaus}, which is the Stokes vector of light, then $H$ describes the polarization rotation by the atoms (Faraday rotation).

In the non-uniformly coupled system of $N$ atoms, the Hamiltonian of Eq.~\ref{eq1} becomes
\begin{equation}
H=\hbar\Omega \sum^N_{j=1}\eta_j \hat s^{(j)}_z \hat A,\label{eq2}
\end{equation}
where $\eta_j$ is the coupling strength of atom $j$ that is proportional to the local light intensity. If we are probing the atoms with a standing-wave beam in an optical resonator on the resonator axis, $\eta_j=\sin^2(k z_j)$ where $z_j$ is the position of the $j$-th atom. If the probing beam is a Gaussian beam in free space or in a running-wave cavity, $\eta_j=\exp[-2(x^2_j+y^2_j)/w^2]$, where $w$ is the beam waist.

\subsection{Special case of uniform atom-light coupling}

We first discuss the case of uniform coupling and then generalize the results to non-uniform coupling. Let us define collective spin operators by
\begin{equation}
\hat S_\alpha=\sum_{j=1}^N \hat s^{(j)}_\alpha.
\end{equation}
Here $\alpha=\{x,y,z\}$. Similarly, we generalize the definition of the raising and lowering spin operators along the $x$ axis by
\begin{eqnarray}
\hat S_{+,k}=\sum_{j=1}^N e^{i {k j 2\pi /N}}\hat s^{(j)}_+,\\
\hat S_{-,k}=\sum_{j=1}^N e^{-i {k j 2\pi /N}}\hat s^{(j)}_-.
\end{eqnarray}
Here, $\hat s^{(j)}_\pm=\hat s^{(j)}_y\pm i \hat s^{(j)}_z$ is the spin raising (lowering) operator for atom $j$ and $k=\{0,1,2,\ldots,N-1\}$. Any atomic state can be decomposed into a combination of different eigenstates of $\vec S^2$ and $\hat S_x$, namely the Dicke states \cite{Dicke}.

Using the Holstein-Primakoff transformation \cite{HP}, when $S\equiv Ns\gg 1$, we treat $|S,S_x=-S\rangle$ as the ground state $|0\rangle$ and write creation and annihilation operators as
\begin{eqnarray}
\hat b_{k}=\hat S_{-,k}/\sqrt{Ns}, \\
\hat b^\dagger_{k}=\hat S_{+,k}/\sqrt{Ns}.
\end{eqnarray}
These operators satisfy the boson commutation relation $[\hat b_k,\hat b^\dagger_l]=\delta_{kl}$. For convenience of notation, we replace $\hat b_0$ ($\hat b^\dagger_0$) by $\hat a$ ($\hat a^\dagger$) in the following.

It is straightforward to verify that the $n$-th excited state
$$(\hat a^\dagger)^n|0\rangle=\sqrt{n!}|S,-S+n\rangle,$$
is the $n$-th Dicke state $\ket{S,-S+n}$ of the atomic ensemble. It is also easy to show that the ground states $\ket{0}_{S-P}$ of the Dicke manifold with total spin $S-P$ \cite{Dicke}, where $P=\sum_{j=1}^{N-1}p_j$, and $p_j$ are integers, can be generalized as $\ket{0}_{S-P}=\ket{S-P,-S+P}=\sqrt{1/\left(\prod_{j=1}^{N-1} p_j!\right)}\prod_{j=1}^{N-1}(\hat b^\dagger_j)^{p_j}\ket{0}$. The corresponding excited Dicke states are given by $(\hat a^\dagger)^n|0\rangle_{S-P}=\sqrt{n!}|S-P, -S+P+n\rangle$. Using this formula, due to $|S_x|\approx S$, we expect the $n$-th excited state of the symmetric manifold $|S-P,-S+P+n\rangle$ to have approximately the same spin distribution probability as $|S, -S+n\rangle$ along any axis in the $y-z$ plane as long as $P\ll S$.

As long as the curvature of the Bloch sphere can be neglected, i.e. $|S_x|\gg1$ and $|S_y|,|S_z|\ll S$, the spin states can be mapped locally onto harmonic oscillator states \cite{CSS}. Then for the state $\ket{S-P,-S+P+n}$ the probability amplitude $g(S_\beta,n)$ to observe a spin $S_\beta$ in the measurement along the axis $S_z \cos(\beta)+\hat S_y \sin(\beta)$ is
\begin{equation}
g(S_\beta,n)={1\over \sqrt{2^n n!}}\left(1\over\pi S\right)^{1/4}e^{i n\beta-S^2_\beta/(2S)}H_n(\sqrt{1\over S}S_\beta),
\end{equation}
where $H_n(x)$ is the $n$-th order Hermite polynomial.

\subsection{The generalization to non-uniform coupling}

Now we generalize the above expressions for the case of non-uniform coupling. For a given Hamiltonian $H=\hbar\Omega \sum^N_{j=1}\eta_j \hat s^{(j)}_z \hat A$, we define the effective spin operators $\tilde S_\alpha={1\over \eta_{eff}}\sum_{j=1}^N \eta_j \hat s^{(j)}_\alpha$, for $\alpha=x$, $y$, $z$.

In order to preserve the commutation relation and the Heisenberg uncertainty principle, we require $\mean{[\tilde S_y,\tilde S_z]}=i\hbar \mean{\tilde S_x}$. Therefore we define the effective coupling $\eta_{eff}$ as \cite{SpinMeasurement}
\begin{equation}
\eta_{eff}={\sum_{j=1}^N \eta^2_j\over \sum_{j=1}^N \eta_j}={\mean{\eta^2}\over\mean{\eta}}.
\end{equation}

The collective creation and annihilation operators are defined in a similar way:
\begin{eqnarray}
\hat b_k=\sum_{j=1}^N f_{k,j} \hat s^{(j)}_-/\sqrt{s}, \\
\hat b^\dagger_k=\sum_{j=1}^N f^*_{k,j} \hat s^{(j)}_+ /\sqrt{s}.
\end{eqnarray}
$f_{k,j}$ satisfies $\sum_{j=1}^N f_{k,j}f^*_{m,j}=\delta_{km}$, and for $\hat b_0$ and $\hat b^\dagger_0$, we choose $f_{0,j}=\eta_j/\sqrt{\sum_{l=1}^N \eta^2_l}$. For the other $f_{k,j}$, we are not interested in the explicit expressions, as linear algebra theory guarantees the existence of these coefficients. $H=\hbar\Omega \sum^N_{j=1}\eta_j \hat s^{(j)}_z \hat A$ commutes with any $\hat b_k$ or $\hat b^\dagger_k$ ($k\ge 1$) for $|S_x|\gg1$, so any initial states will remain on the same Bloch sphere under the action of the Hamiltonian.

In this situation, we can define effective Dicke states which have the same observable properties as the Dicke states under uniform coupling.
\begin{equation}
{(\hat a^\dagger)^n\prod_{j=1}^{N-1} (\hat b^\dagger_k)^{p_j}|0\rangle\over\sqrt{n!\prod_{j=1}^{N-1}p_j!}}=|S_e-P,-S_e+P+n\rangle,
\end{equation}
where $S_e$ is the effective total spin which will be defined later.

Using the interaction described by the Hamiltonian in Eq \ref{eq2}, measurement of the observable $\hat A$ yields information about $\tilde S_z$. By applying an externally driven rotation $\beta$ along $\hat S_x$, one can measure $\tilde S_z\cos(\beta)+\tilde S_y\sin(\beta)$, and we label this measurement as $\tilde S_\beta$. As long as the resolution is not high enough to distinguish each individual spin, which is common in an atomic ensemble, the collective spin can be treated as a continuous variable \cite{Klaus}. So the probability amplitude to measure a particular $\tilde S_\beta$ is
\begin{equation}
g(\tilde S_\beta,n)={1\over \sqrt{2^n n!}}\left(1\over\pi N_{e}s\right)^{1/4}e^{i n\beta-\tilde S^2_\beta/(2N_{e} s)}H_n(\sqrt{1\over N_{e}s}\tilde S_\beta).
\end{equation}
Here, $N_{e}=N \mean{\eta}/\eta_{eff}$, is the effective atom number, and $S_e=N_e s$ is the effective total spin. The idea of an effective atom number was first introduced in Refs \cite{SpinDy,SpinMeasurement} for characterizing Gaussian spin distribution, and we have derived it here more generally from the Heisenberg uncertainty.

Therefore, by using effective operators and an effective atom number, the physical observables remain the same as under uniform coupling. The equivalence also applies to any atomic states satisfying $|\tilde S_x|\approx N_e s\gg 1$. The Hamiltonian is simply written as $H=\hbar\tilde\Omega\tilde S_z\hat A$ where $\tilde\Omega=\eta_{eff}\Omega$. Then, all the predictions for the non-uniform coupling are equivalent to those for uniform coupling.

\begin{figure}[htbp]
\begin{center}
\includegraphics[height=6in]{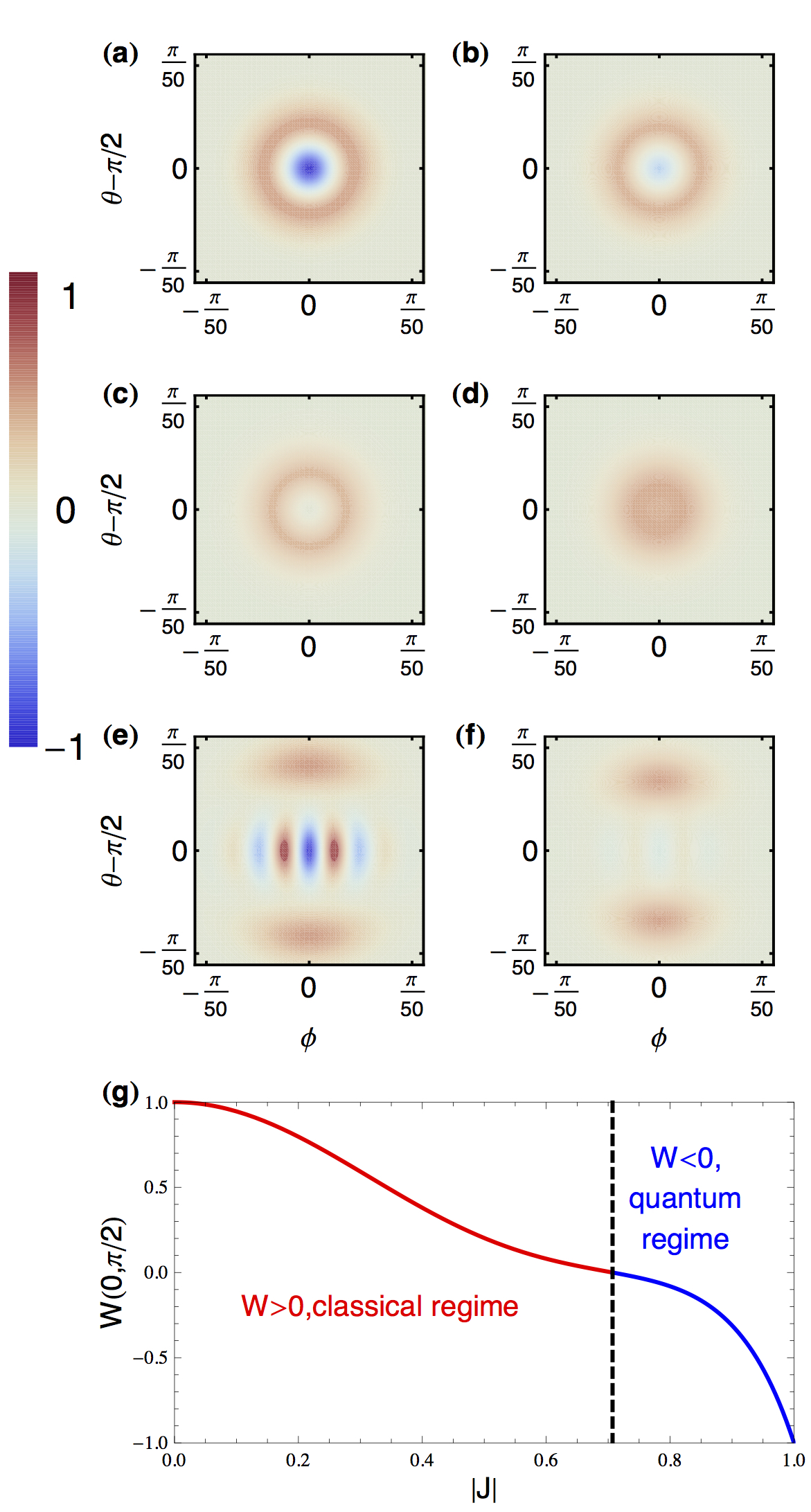}
\caption{(a)-(d) The Wigner function \cite{Wigner} for the first Dicke state prepared and detected with different mode functions, for overlap parameter $J=1,\sqrt{2/3},\sqrt{1/2}$ and $\sqrt{3/10}$. (a) The center of Wigner function $W(0,\pi/2)$ reaches $-1$, the maximal allowed negative value, when we use the same non-uniform coupling for state preparation and readout, independent of the choice of the coupling. (b) If the preperation mode is a standing wave and the readout mode has uniform coupling, then $J=\sqrt{2/3}$, and $W(0,\pi/2)$ is only $-1/3$. (c)-(d) When $J$ continues decreasing, the negative value of the Wigner function is smeared out due to the mismatch between state generation and readout. (e) For a squeezed cat state generated by a five-photons heralding event \cite{PRAtheory}, the interference fringes still maintain the maximal visibility when $J=1$, regardless of the coupling of each atom. (f) The fringes are smeared out when $J$ decreases (here $J=\sqrt{2/3}$). (g) The dependence of $W(0,\pi/2)$ on $|J|$ for both states. All graphs are shown for atomic spin $s=1$ and $N=2000$ atoms.}
\label{fig2}
\end{center}
\end{figure}

\subsection{Connecting different coupling modes}

Based on the analysis above, we can also derive a useful formula connecting different effective Dicke states. If we have two Hamiltonians with different atom-light coupling $H=\hbar\Omega \sum^N_{j=1}\eta_j \hat s^{(j)}_z \hat A$ and $H'=\hbar\Omega \sum^N_{j=1}\xi_j \hat s^{(j)}_z \hat B$, then we can define two sets of creation and annihilation operators $\{\hat b_k,\hat b^\dagger_k\}$ and $\{\hat d_k,\hat d^\dagger_k\}$ as above. For $k=0$, we still have
\begin{eqnarray}
\hat a_\eta=\hat b_0=\sum_{j=1}^N {\eta_j\over\sqrt{s\sum_{k=1}^N\eta^2_k}}\hat s^{(j)}_-,\\
\hat a_\xi=\hat d_0=\sum_{j=1}^N {\xi_j\over\sqrt{s\sum_{k=1}^N\xi^2_k}}\hat s^{(j)}_-.
\end{eqnarray}
We define the overlap parameter $J$ between these two couplings as
\begin{equation}
J={\sum_{l=1}^N\eta_l\xi_l\over \sqrt{\left(\sum_{l=1}^N\eta^2_l\right)\left(\sum_{l=1}^N\xi^2_l\right)}}.
\end{equation}
Without losing generality, we can choose the coefficients $f_{k,j}$ of the set $\{\hat d_k\}$ such that $\hat a_\eta=J\hat a_\xi+\sqrt{1-J^2}\hat d_1$.

Now we consider the state that is prepared on the maximal Bloch sphere of $S_{e,\eta}=N_{e,\eta}s$. Any effective Dicke state $|S_{e,\eta},-S_{e,\eta}+n\rangle_\eta$ on this sphere can be expanded as
\begin{eqnarray}
& &|S_{e,\eta},-S_{e,\eta}+n\rangle_\eta\nonumber\\
&=&{(\hat a^\dagger_\eta)^n\over\sqrt{n!}}|0\rangle
={(J\hat a^\dagger_\xi+\sqrt{1- J^2} d_1^\dagger)^n\over\sqrt{n!}}|0\rangle \\
&=&\sum_{k=0}^n \sqrt{\binom{n}{k}}  J^{n-k}(1- J^2)^{k/2}|S_{e,\xi}-k,-S_{e,\xi}+n\rangle_\xi.\nonumber
\end{eqnarray}

Applying this formula, we need to know just one parameter $J$ to establish the connection between effective Dicke states for different non-uniform coupling bases.

In Fig \ref{fig2}, we show a few examples illustrating the effects of non-uniform coupling. If we prepare and probe the first Dicke state with the same non uniform coupling, the Wigner function distribution reaches -1, the most non-classical value, and is identical to the Wigner function for uniform coupling. However, if we were to measure the same state in another coupling basis, we would find a reduced value for the magnitude of the negative Wigner function at the origin. If the overlap parameter $J$ is decreased further, the central hole $W(0,\pi/2)$ in the Wigner function $W$ will be smeared out by the growing mismatch between the couplings used for state preparation and observation, respectively. Moreover, this effect is more obvious for a cat state in Fig, \ref{fig2}(e), (f), since the narrower fringes are more fragile than a wide hole. In fact, there is a general relation for any quantum state. When $|J|$ is below $0.71$, the Wigner function is all positive, which corresponds to a classical probability distribution. Quantum interference with $W<0$ can only be seen when $|J|>0.71$.

Another interesting example is the squeezed state. If the preparation and readout couplings are identical, the non-uniform coupling will not affect the squeezing parameter. The theoretical prediction and analysis under uniform coupling are still valid. The only correction needed is to replace the atom number $N$ by the effective atom number $N_e$ \cite{SpinMeasurement,SpinDy}.

\begin{figure}
  \includegraphics[width=2.8in]{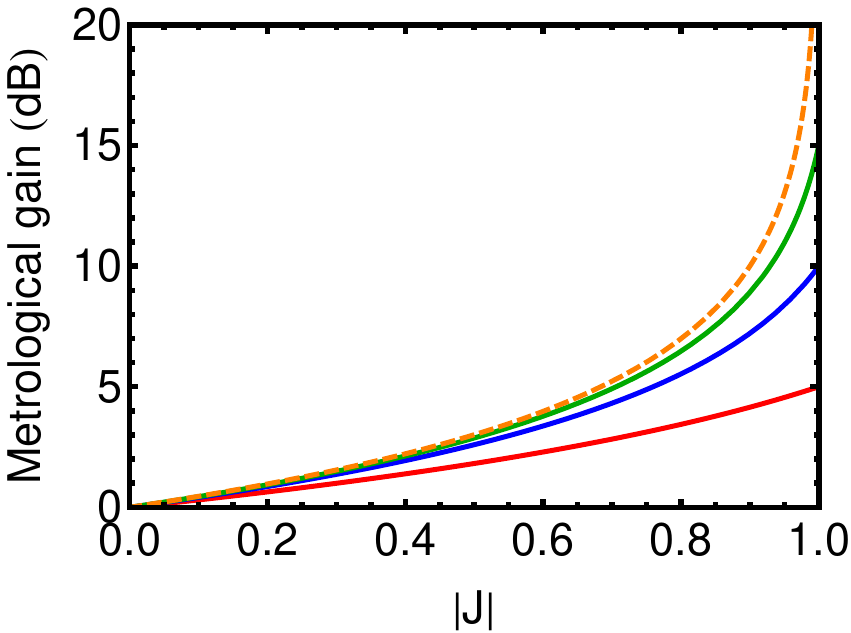}\\
  \caption{Spin squeezing using different couplings for state preparation and measurement. The solid lines show the metrological gain versus $J$ for spin squeezed states with 15dB, 10dB and 5dB of squeezing (from above to below). The dashed line is the upper bound $G$ for any squeezed state, where $G=1/\left[(1-J^2)/2+1/ S\right]$. For $J<1$, the mismatch limits the maximum possible metrological gain.}\label{fig3}
\end{figure}

If there are different couplings involved in generating and observing in the squeezed state, the squeezing parameter will decrease when $|J|<1$. In this case, the observable squeezing and the metrological gain are limited by the coupling overlap $J$. For any given $J$, there is always an upper bound of the metrological gain for any squeezed state. The results are summarized in Fig \ref{fig3}. This limitation could be important, e.g., for the operation of spin squeezed atom interferometers, where the atoms may be detected in a different position than where they were prepared.

Consider a squeezed state that approaches the Heisenberg limit, $\Delta S^2_z/(Ns)^2\sim1/(Ns)^2$. The scaling of variance becomes $1/N^2$ instead of $1/N$. However, the mismatch, due to effects such as atomic thermal motion, limits the detectable squeezing when we use this state in a precision measurement. We find that when $1-|J|>>1/N$, the best observable squeezing during the readout deteriorates to $\Delta \tilde S^2_z/(S)^2=1/(S)^2+(1-J^2)/(2S)$. The variance now scales again as $1/N$, not $1/N^2$. This shows that any change in the atom-light coupling between state preparation and readout larger than $1/N$ will destroy the Heisenberg-limited scaling. We use atoms trapped in an optical cavity as an example to illustrate the effect of finite temperature. We assume that the dipole trap used to confine the atoms and the probing light field have the same spatial mode. The thermal random motion reduces the parameter $J$ as $1-|J|\approx (k_B T/U)^2$, where $T$ is the temperature and $U$ is the trap depth. In order to observe the Heisenberg limit, the required temperature $T$ is below $U/(k_B \sqrt N)$. For a trap depth of 10MHz and $N=10^8$ atoms, the ensemble must be cooled down to 100~nK to reach the Heisenberg limit.

\section{Conclusion}

In conclusion, we have shown the equivalence between uniform coupling and non-uniform coupling in the optical preparation and detection of collective atomic spin states as long as no measurements with single-atom resolution are performed. This eliminates some conceptional concerns about entanglement in real, non-uniformly coupled systems. By using the effective spin and atom number, the collective evolution of the system can be described and predicted. We also derive a useful formula that can be used to calculate, e.g. the observable squeezing at finite atomic temperature or when an entangled atomic state is prepared with a different light mode than used for detection, e.g. in an atom interferometer \cite{PhysRevLett.98.111102}.

\section{Acknowledgement}

This work was supported by the NSF, DARPA (QUASAR), and MURI grants through AFOSR and ARO.
\bibliographystyle{apsrev4-1}
\bibliography{NonUniform0921}

\end{document}